\documentclass[journal,10pt]{IEEEtran}
\IEEEoverridecommandlockouts
\usepackage[T1]{fontenc}
\usepackage{cite}
\usepackage{amsmath,amssymb,amsfonts}
\usepackage[ruled,vlined]{algorithm2e}
\usepackage{graphicx}
\usepackage{textcomp}
\usepackage{xcolor}
\usepackage{booktabs} 
\usepackage{multirow}
\usepackage{diagbox}
\usepackage{url}
\usepackage{threeparttable}
\usepackage{array}
\usepackage{textcase}
\usepackage{rotating}
\usepackage{etoolbox}
\makeatletter
\let\MYcaption\@makecaption
\makeatother

\usepackage[font=footnotesize]{subcaption}

\makeatletter
\let\@makecaption\MYcaption
\makeatother

\makeatletter
\def\UrlAlphabet{%
      \do\a\do\b\do\c\do\d\do\e\do\f\do\g\do\h\do\i\do\j%
      \do\k\do\l\do\m\do\n\do\o\do\p\do\q\do\r\do\s\do\t%
      \do\u\do\v\do\w\do\x\do\y\do\z\do\A\do\B\do\C\do\D%
      \do\E\do\F\do\G\do\H\do\I\do\J\do\K\do\L\do\M\do\N%
      \do\O\do\P\do\Q\do\R\do\S\do\T\do\U\do\V\do\W\do\X%
      \do\Y\do\Z}
\def\UrlDigits{\do\1\do\2\do\3\do\4\do\5\do\6\do\7\do\8\do\9\do\0}
\g@addto@macro{\UrlBreaks}{\UrlOrds}
\g@addto@macro{\UrlBreaks}{\UrlAlphabet}
\g@addto@macro{\UrlBreaks}{\UrlDigits}

\patchcmd{\@algocf@start}
  {-1.5em}
  {0pt}
  {}{}

\makeatother

\graphicspath{{figures/}}

\begin{document}
\ifdefined \GramaCheck
  \newcommand{\CheckRmv}[1]{}
  \newcommand{\figref}[1]{Figure 1}%
  \newcommand{\tabref}[1]{Table 1}%
  \newcommand{\secref}[1]{Section 1}
  \newcommand{\algref}[1]{Algorithm 1}
  \renewcommand{\eqref}[1]{Equation 1}
\else
  \newcommand{\CheckRmv}[1]{#1}
  \newcommand{\figref}[1]{Fig.~\ref{#1}}%
  \newcommand{\tabref}[1]{Table~\ref{#1}}%
  \newcommand{\secref}[1]{Sec.~\ref{#1}}
  \newcommand{\algref}[1]{Algorithm~\ref{#1}}
  \renewcommand{\eqref}[1]{(\ref{#1})}
\fi
\newtheorem{theorem}{Theorem}
\newtheorem{proposition}{Proposition}
\newtheorem{assumption}{Assumption}
\newtheorem{definition}{Definition}
\newtheorem{condition}{Condition}
\newtheorem{property}{Property}
\newtheorem{remark}{Remark}
\newtheorem{lemma}{Lemma}
\newtheorem{corollary}{Corollary}
%
\title{
Robust MIMO Channel Estimation Using Energy-Based Generative Diffusion Models
}

\author{Ziqi~Diao,
        Xingyu~Zhou,
        Le~Liang,~\IEEEmembership{Member,~IEEE,}
        and~Shi~Jin,~\IEEEmembership{Fellow,~IEEE}
\thanks{Z.~Diao, X.~Zhou, L.~Liang, and S.~Jin are with the School of Information Science and Engineering, Southeast University, Nanjing 210096, China
(e-mail: \protect \url{ziqidiao@seu.edu.cn}; \protect \url{xy_zhou@seu.edu.cn}; lliang@seu.edu.cn; jinshi@seu.edu.cn). L. Liang is also with Purple Mountain Laboratories, Nanjing 211111, China.}
}
%
%

\maketitle

\begin{abstract}
  Channel estimation for massive multiple-input multiple-output (MIMO) systems is fundamentally constrained by excessive pilot overhead and high estimation latency.
  To overcome these obstacles, recent studies have leveraged deep generative networks to capture the prior distribution of wireless channels. 
  In this paper, we propose a novel estimation framework that integrates an energy-based generative diffusion model (DM) with the Metropolis-Hastings (MH) principle.
  By reparameterizing the diffusion process with an incorporated energy function, the framework explicitly estimates the unnormalized log-prior, while MH corrections refine the sampling trajectory, mitigate deviations, and enhance robustness, ultimately enabling accurate posterior sampling for high-fidelity channel estimation.
  Numerical results reveal that the proposed approach significantly improves estimation accuracy compared with conventional parameterized DMs and other baseline methods, particularly in cases with limited pilot overhead.

\end{abstract}


\vspace{-0.1cm}
\section{Introduction}


Massive multiple-input multiple-output (MIMO) is a key enabler for next-generation wireless communication systems, offering substantial gains in both spectral and energy efficiency.
The performance of massive MIMO systems critically depends on accurate channel estimation, as the channel state information (CSI) directly impacts precoding, coherent detection, and resource allocation \cite{luoverviewmassive2014}. 
However, the large antenna arrays significantly increase estimation complexity, making reliable acquisition of high-dimensional CSI particularly challenging under limited pilot overhead.



To address these challenges, recent researches have explored the use of generative models, which have been applied to various wireless communication tasks \cite{celikdawagenerative2024}. 
For channel estimation, several generative learning approaches have been leveraged, such as generative adversarial networks, Gaussian mixture models, and variational autoencoders. 
Among these, diffusion models (DMs) have emerged as the most prevailing approach and have been widely adopted to tackle the challenges of high-dimensional channel estimation \cite{arvintescore-based2023,fesllowcommplexity2024,zhoudm2025,zilberstein2024joint,chen2025generative}.

In \cite{arvintescore-based2023}, a score-based generative model (SGM) was proposed to address channel estimation through posterior sampling. 
Building upon this, a novel variational inference approach was developed in \cite{chen2025generative} for channel estimation by utilizing a pre-trained SGM. 
Both approaches exhibit robust estimation performance, achieving superior accuracy in both in-distribution and out-of-distribution scenarios. 
The study in \cite{fesllowcommplexity2024} designed a DM with a lightweight convolutional neural network (CNN) backbone to learn the channel distribution in the angular domain, demonstrating outstanding performance across different signal-to-noise ratios (SNRs). 
Moreover, the DM-based estimator was extended to quantized and under-determined massive MIMO systems in \cite{zhoudm2025}.
In \cite{zilberstein2024joint}, the discrete prior of the transmitted symbols and a learned channel prior were incorporated into SGM to address the joint massive MIMO channel estimation and data detection problem. 

Despite these advancements, existing studies are still limited by either excessive computational complexity or inadequate estimation accuracy. 
For instance, certain SGM-based approaches are constrained by a large number of parameters and substantial reverse steps, which hinder real-time estimation.
In addition, some methods rely on orthogonal and full pilot measurements, restricting their applicability in practical scenarios.
Moreover, existing DM-based approaches rely primarily on unadjusted samplers.
Such reliance often leads to distributional mismatches between the generated samples and the target, which in turn degrade the performance of posterior sampling. These limitations underscore the necessity of developing more precise and robust samplers.

In this paper, we propose a massive MIMO channel estimation method based on an energy-based DM. 
We introduce an energy function to explicitly model the channel's log-prior distribution and learn it in an unsupervised manner. 
Building upon this, we develop a posterior sampling process enhanced with Metropolis-Hastings (MH) corrections \cite{roberts2002langevin}. 
The proposed method achieves notably enhanced CSI acquisition accuracy compared to state-of-the-art approaches, particularly in the low-pilot-overhead regime, while maintaining low complexity.

\section{System Model and Preliminaries}
\label{sec:sysmodel}
\subsection{MIMO Channel Estimation}  
We consider MIMO channel estimation in a narrowband, point-to-point communication scenario with $N_{\mathrm{t}}$ transmit and $N_{\mathrm{r}}$ receive antennas. Extension to a broadband orthogonal frequency division multiplexing (OFDM) system is straightforward. 
Both the transmitter and the receiver are equipped with uniform linear arrays with half-wavelength spacing.
A total of $N_{\mathrm{p}}$ pilot symbols are allocated, and we assume that the channel remains unchanged during pilot transmissions. 
The received signal $\mathbf{Y}\in \mathbb{C}^{N_{\mathrm{r}} \times N_{\mathrm{p}}}$ can be expressed as
\CheckRmv{
  \begin{equation}
     \mathbf{Y}=\mathbf{HP}+\mathbf{N},
     \label{eq:model1}
  \end{equation}
}
where $\mathbf{H}\in \mathbb{C}^{N_{\mathrm{r}} \times N_{\mathrm{t}}}$ represents the complex-valued channel matrix to be estimated, $\mathbf{P}\in \mathbb{C}^{N_{\mathrm{t}} \times N_{\mathrm{p}}}$ is the pilot matrix, and $\mathbf{N}\in \mathbb{C}^{N_{\mathrm{r}} \times N_{\mathrm{p}}}$ is the complex additive white Gaussian noise (AWGN) with each entry being independently and identically distributed with mean zero and variance $2\sigma^2$. 
Each entry in $\mathbf{P}$ is a randomly chosen unit-power quadrature phase shift keying (QPSK) symbol.
Using vectorization, \eqref{eq:model1} can be rewritten as
\CheckRmv{
  \begin{equation}
     \mathbf{y}_\mathrm{c}=\mathbf{A}_\mathrm{c}\mathbf{h}_\mathrm{c}+\mathbf{n}_\mathrm{c},
     \label{eq:model2}
  \end{equation}
}
where $\mathbf{A}_\mathrm{c}\in \mathbb{C}^{N_{\mathrm{r}}N_{\mathrm{p}} \times {N_{\mathrm{r}}N_{\mathrm{t}}}}$ is the measurement matrix given by $\mathbf{A}_\mathrm{c}=\mathbf{P}^T\otimes\mathbf{I}_{N_\mathrm{r}}$,
with $\otimes$ denoting the Kronecker product and $\mathbf{I}$ denoting an identity matrix. 
We normalize the channel as $\mathbb{E}[|h_{ij}|^2] = 1$ and define the SNR as $N_{\mathrm{t}}/(2\sigma^2)$.
Furthermore, the vectorized channel $\mathbf{h}_\mathrm{c}$ is related to its angular domain representation via the discrete Fourier transform (DFT) as \cite{bajwa2010}
\CheckRmv{  
  \begin{equation}
    \mathbf{h}_\mathrm{c} = \left(\mathbf{F}_\mathrm{T}^{\mathrm{H}} \otimes \mathbf{F}_\mathrm{R}\right)\mathbf{h}_\mathrm{ad},
    \label{eq:angular_channel}
  \end{equation}
}
where $\mathbf{h}_\mathrm{ad}$ represents the corresponding angular domain channel, $\mathbf{F}_\mathrm{T} \in \mathbb{C}^{N_{\mathrm{t}} \times N_{\mathrm{t}}}$ and $\mathbf{F}_\mathrm{R} \in \mathbb{C}^{N_{\mathrm{r}} \times N_{\mathrm{r}}}$ denote the DFT matrices associated with the transmit and receive antenna arrays, respectively.
Combining \eqref{eq:model2} and \eqref{eq:angular_channel}, we have
\CheckRmv{
  \begin{equation}
    \mathbf{y}_\mathrm{c}=\mathbf{A}_{\mathrm{ad}}{\mathbf{h}}_{\mathrm{ad}}+\mathbf{n}_\mathrm{c},
    \label{eq:model4}
  \end{equation}
}
where $\mathbf{A}_{\mathrm{ad}}= \mathbf{A}_\mathrm{c}\left(\mathbf{F}_\mathrm{T}^{\mathrm{H}} \otimes \mathbf{F}_\mathrm{R}\right) \in\mathbb{C}^{N_\mathrm{r}N_\mathrm{p}\times{N_\mathrm{r}N_\mathrm{t}}}$.
This model is further transformed to the real-valued form written as
\CheckRmv{
  \begin{equation}
     \mathbf{y}=\mathbf{Ah}+\mathbf{n},
     \label{eq:model3}
  \end{equation}
}
where $\mathbf{y},\mathbf{n}\in\mathbb{R}^{M\times1}$, $\mathbf{A}\in\mathbb{R}^{M\times N}$, and $\mathbf{h}\in\mathbb{R}^{N\times1}$, with the dimensions specified as $M=2N_\mathrm{r}N_\mathrm{p}$ and $N=2N_\mathrm{r}N_\mathrm{t}$. 
The noise vector $\mathbf{n}\sim\mathcal{N}(\mathbf{0},\sigma^2\mathbf{I}_M)$, where $\mathcal{N}(\boldsymbol{\mu},\boldsymbol{\Omega})$ represents a Gaussian distribution with mean $\boldsymbol{\mu}$ and covariance $\boldsymbol{\Omega}$.
The MIMO channel estimation problem formulated in \eqref{eq:model3} is to recover the angular domain channel vector $\mathbf{h}$, provided the triplet \{$\mathbf{y}, \mathbf{A}, \sigma^2$\}.
Let $\alpha = N_{\mathrm{p}}/N_{\mathrm{t}}$ denote the pilot density.
For the case $\alpha < 1$, channel estimation corresponds to an under-determined linear inverse problem, which considerably complicates CSI recovery.

\subsection{Diffusion Model for Channel Estimation}
\label{sec:twob}
DMs have shown outstanding ability in modeling complex distributions, which bring new ideas and approaches to channel estimation.
Given a data distribution $\mathbf{h}_0\sim p_0(\cdot)$, the forward process of the DM adds noise using a Gaussian transition kernel for $T$ time steps. Specifically, each transition is defined as $p(\mathbf{h}_t|\mathbf{h}_{t-1})=\mathcal{N}(\mathbf{h}_t;\sqrt{1-\beta_t}\mathbf{h}_{t-1},\beta_t\mathbf{I})$,
where $\mathbf{h}_t$ is the latent variable at time step $t$, $0<\beta_1<\beta_2<\ldots<\beta_T<1$ denote the noise schedule. 
Defining $\alpha_{t}\triangleq1-\beta_{t}$ and $\bar{\alpha}_{t}\triangleq\prod_{i=1}^{t}\alpha_{i}$ and
by iteratively reparameterizing, we obtain
\CheckRmv{
  \begin{equation}
     \mathbf{h}_t=\sqrt{\bar{\alpha}_t}\mathbf{h}_0+\sqrt{1-\bar{\alpha}_t}\boldsymbol{\epsilon}_t,\quad \boldsymbol{\epsilon}_t\sim\mathcal{N}(\mathbf{0},\mathbf{I}).
     \label{eq:forward} 
  \end{equation}
}
The noise schedule is prescribed such that $\mathbf{h}_T$ approximately follows $\mathcal{N}(\mathbf{0},\mathbf{I})$.

The reverse process of the DM takes the form $ p_{\boldsymbol{\theta}}(\mathbf{h}_{t-1}|\mathbf{h}_t)$ parameterized by $\boldsymbol{\theta}$ to learn the reverse distribution $p(\mathbf{h}_{t-1}|\mathbf{h}_t)$ which denoises $\mathbf{h}_t$ to $\mathbf{h}_{t-1}$.
Note $p(\mathbf{h}_{t-1}|\mathbf{h}_t)$ is generally intractable due to the unknown prior $ p_0(\cdot)$.
However, it becomes tractable when conditioned on $\mathbf{h}_0$, i.e., $p(\mathbf{h}_{t-1}|\mathbf{h}_t,\mathbf{h}_0)=\mathcal{N}(\mathbf{h}_{t-1};\tilde{\boldsymbol{\mu}}_t,\tilde{\beta}_t\mathbf{I})$,
where $\tilde{\boldsymbol{\mu}}_{t} =\frac{1}{\sqrt{\alpha_t}}\left(\mathbf{h}_t-\frac{1-\alpha_t}{\sqrt{1-\bar{\alpha}_t}}\boldsymbol{\epsilon}_t\right)$, 
and $\tilde{\beta}_{t} =\frac{1-\bar{\alpha}_{t-1}}{1-\bar{\alpha}_{t}}\beta_{t}$.
Hence, the model $p_{\boldsymbol{\theta}}(\mathbf{h}_{t-1}|\mathbf{h}_t)$ is parameterized as Gaussian, given by $\mathcal{N}(\mathbf{h}_{t-1};\boldsymbol{\mu_{\boldsymbol{\theta}}}(\mathbf{h}_{t},t),\tilde{\beta}_{t}\mathbf{I})$, 
where $\boldsymbol{\mu_{\boldsymbol{\theta}}}$ serves to approximate $\tilde{\boldsymbol{\mu}}_{t}$. 
Typically, we use the $\boldsymbol{\epsilon}_{\boldsymbol{\theta}}$-parameterization, i.e., replacing $\boldsymbol{\epsilon}_t$ with a neural network $\boldsymbol{\epsilon_{\boldsymbol{\theta}}}(\mathbf{h}_t,t)$,
which is denoted as the denoising network.
In this case, 
the mean of the reverse process
can be expressed as
\CheckRmv{
  \begin{equation}
   \boldsymbol{\mu_{\boldsymbol{\theta}}}(\mathbf{h}_t,t)=\frac{1}{\sqrt{\alpha_t}}\left(\mathbf{h}_t-\frac{1-\alpha_t}{\sqrt{1-\bar{\alpha}_t}}\boldsymbol{\epsilon}_{{\boldsymbol{\theta}}}(\mathbf{h}_t,t)\right),
    \label{eq:reverse}
  \end{equation}
}
and the DM can be trained with the loss
\CheckRmv{
  \begin{equation}
    \mathcal{L}_{\mathrm{DM}}(\boldsymbol{\theta})=\mathbb{E}_{\mathbf{h}_0,\boldsymbol{\epsilon}_t,t}\left[\left\|\boldsymbol{\epsilon}_t-\boldsymbol{\epsilon}_{\boldsymbol{\theta}}(\sqrt{\bar{\alpha}_t}\mathbf{h}_0+\sqrt{1-\bar{\alpha}_t}\boldsymbol{\epsilon}_t,t)\right\|_2^2\right],
    \label{eq:dmloss}
  \end{equation}
}
where $\left\|\cdot\right\|_2$ denotes the $l_2$-norm.
After training, the parameters of $\boldsymbol{\epsilon}_{\boldsymbol{\theta}}$ are fixed, and data samples are iteratively generated for $t=T$ to 1 using
\CheckRmv{
  \begin{equation}
    \mathbf{h}_{t-1}=\frac{1}{\sqrt{\alpha_t}}\left(\mathbf{h}_t-\frac{1-\alpha_t}{\sqrt{1-\bar{\alpha}_t}}\boldsymbol{\epsilon_{\boldsymbol{\theta}}}(\mathbf{h}_t,t)\right)+\tilde{\beta}_t\mathbf{z}_t,
    \label{eq:sample}
  \end{equation}
}
where $\mathbf{z}_t\sim\mathcal{N}(\mathbf{0},\mathbf{I})$.

As discussed in \cite{arvintescore-based2023,zhoudm2025}, the channel can be estimated via iterative posterior sampling,
where each time step $t$ (from $t = T$ to 1) gives a new proposal sample \cite{zhoudm2025}
\CheckRmv{
  \begin{equation}
    \mathbf{h}_{\mathrm{prop}}=\frac{1}{\sqrt{\alpha}_t}\left[\mathbf{h}_t+(1-\alpha_t)\nabla_{\mathbf{h}_t}\log p(\mathbf{h}_t|\mathbf{y})\right]+\tilde{\beta}_t\mathbf{z}_t,
    \label{eq:proposal}
  \end{equation}
}
where the gradient of the log-posterior $\nabla_{\mathbf{h}_t}\log p(\mathbf{h}_t|\mathbf{y})$, namely posterior score, can be decomposed into the prior score and likelihood score utilizing the Bayes' rule: $\nabla_{\mathbf{h}_t}\log p(\mathbf{h}_t|\mathbf{y}) \! = \! \nabla_{\mathbf{h}_t}\log p(\mathbf{h}_t) \! + \! \nabla_{\mathbf{h}_t}\log p(\mathbf{y}|\mathbf{h}_t)$.
Then, the prior score can be approximated using the  $\boldsymbol{\epsilon}_{\boldsymbol{\theta}}$-parameterized DM \cite{zhoudm2025}, i.e., the pre-trained denoising network $\boldsymbol{\epsilon}_{\boldsymbol{\theta}}$:
\CheckRmv{
  \begin{equation}
    \nabla_{\mathbf{h}_t}\log p(\mathbf{h}_t)\approx-\frac{1}{\sqrt{1-\bar{\alpha}_t}}\boldsymbol{\epsilon}_{\boldsymbol{\theta}}(\mathbf{h}_t,t).
    \label{eq:dm}
  \end{equation}
}
Meanwhile, the likelihood score can be approximated using \cite[Eq. (22)]{zhoudm2025}. In \cite{arvintescore-based2023,zhoudm2025}, the proposal in \eqref{eq:proposal} is directly accepted as $\mathbf{h}_{t-1}$, and the final estimate $\hat{\mathbf{h}}$ is derived after $T$ iterations of \eqref{eq:proposal}.
\section{Proposed Method} %

\subsection{Energy-Based Diffusion Model}


Using the $\boldsymbol{\epsilon}_{\boldsymbol{\theta}}$-parameterization introduced above, only the score can be estimated
and all new proposals given in \eqref{eq:proposal} are simply accepted.
This scheme, typically termed the unadjusted Langevin algorithm, may generate samples that significantly deviate from the target posterior distribution, thereby leading to performance degradation \cite{zhoudm2025}. 
Consequently, the development of more accurate sampling is essential to enhance estimation performance.

MH corrections have demonstrated promise in improving sample quality and convergence \cite{roberts2002langevin,zhou2024near}. 
This method performs an acceptance test on the proposals, i.e., accepting or rejecting them based on an acceptance rate.
However, the incorporation of these corrections into DMs is non-trivial due to the lack of access to the log-prior $\log p(\mathbf{h})$ under $\boldsymbol{\epsilon}_{\boldsymbol{\theta}}$-parameterization, which is necessary for computing the MH acceptance rate.
To overcome this limitation, a change in the parameterization of DMs is proposed to enable the use of MH corrections, which allows us to refine the sampling process and correct for deviations at each time step.


The key component of our approach is an energy-based parameterization of the DM, briefly termed energy-based DM \cite{re32023}.
Instead of directly using the denoising neural network $\boldsymbol{\epsilon}_{\boldsymbol{\theta}}(\mathbf{h},t)$ to parameterize the DM as introduced in \secref{sec:twob}, we define a scalar-output neural network $f_{\boldsymbol{\theta}}(\mathbf{h},t)$, also referred to as the energy function.
The proposed energy-based DM parameterizes a distribution as
\CheckRmv{
  \begin{equation}
    p_{\boldsymbol{\theta}}(\mathbf{h},t)=\frac{1}{Z(\boldsymbol{\theta})}e^{f_{\boldsymbol{\theta}}(\mathbf{h},t)}, 
  \end{equation}
}
where $Z(\boldsymbol{\theta})= \int e^{f_{\boldsymbol{\theta}}(\mathbf{h},t)}\mathrm{d}\mathbf{h}$ is the normalizing constant for the modeled distribution. 
We then implicitly define $\boldsymbol{\epsilon}_{\boldsymbol{\theta}}$ in \eqref{eq:dm} as the gradient of the energy function:
\CheckRmv{
  \begin{equation}
    \boldsymbol{\epsilon}_{\boldsymbol{\theta}}(\mathbf{h},t)\triangleq\nabla_{\mathbf{h}}f_{\boldsymbol{\theta}}(\mathbf{h},t).
    \label{eq:energyfunc}
  \end{equation}
}
This implicitly-defined $ \boldsymbol{\epsilon}_{\boldsymbol{\theta}}$ is utilized in a manner consistent with the standard DM training paradigm as shown in \eqref{eq:dmloss}.
By combining \eqref{eq:dm} and \eqref{eq:energyfunc}, the log-prior probability can be derived as
\CheckRmv{
  \begin{equation}
    \log p(\mathbf{h}_t)\approx-\frac{1}{\sqrt{1-\bar{\alpha}_t}}f_{\boldsymbol{\theta}}(\mathbf{h}_t,t)+ C,
    \label{eq:log=prior}
  \end{equation}
}
where $C$ is a constant.
In general, leveraging the proposed energy-based DM, we can not only recover $\nabla_{\mathbf{h}_t}\log p(\mathbf{h}_t)$ as before, but also explicitly estimate the unnormalized log-prior probability $\log p(\mathbf{h}_t)$.
This estimation, given in \eqref{eq:log=prior}, plays a crucial role in calculating the MH acceptance rate.

Inspired by the denoising autoencoder \cite{re32023}, we parameterize the energy function as follows to facilitate learning
\CheckRmv{
  \begin{equation}
    f_{\boldsymbol{\theta}}(\mathbf{h},t)=-\frac{1}{2}\left\|\mathbf{h}-\boldsymbol{d}_{\boldsymbol{\theta}}(\mathbf{h},t)\right\|_2^2,
    \label{eq:energy}
  \end{equation}
}
where $\boldsymbol{d}_{\boldsymbol{\theta}}(\mathbf{h},t)$ is a neural network that can adopt the standard denoising network architecture.
Following \cite{fesllowcommplexity2024}, a lightweight CNN architecture, as shown in the upper right part of \figref{fig:ebm_block}, is employed as the backbone of $\boldsymbol{d}_{\boldsymbol{\theta}}$, with trainable parameters $\boldsymbol{\theta}$.
The proposed energy-based DM is trained in an unsupervised manner, as illustrated below.

\subsection{Training Process}
\figref{fig:ebm_block} presents the block diagram of the proposed schemes, with the upper left part showing the forward diffusion in the training phase.
At each training step, a data sample $\mathbf{h}_0$ is randomly drawn from the channel dataset. 
Concurrently, we generate random noise $\boldsymbol{\epsilon}\sim\mathcal{N}(\mathbf{0},\mathbf{I})$ and add it to the selected sample $\mathbf{h}_0$ according to \eqref{eq:forward},
where the ratio $\bar{\alpha}_t$ is determined by a time step index $t$ uniformly chosen from $1,2,\ldots,T$.
Then, the noisy channel data $\mathbf{h}_t$, along with the time step index $t$, are input into the denoising network.
The network's output is utilized to compute the energy value, as defined in \eqref{eq:energy}.
Similar to \eqref{eq:dmloss}, we train the proposed energy-based DM by minimizing the mini-batch version of the following loss function
\CheckRmv{  
  \begin{equation}
    \mathcal{L}_{\mathrm{EBM}}\left(\boldsymbol{\theta}\right)=\mathbb{E}_{\mathbf{h}_t}\big[\left\|\boldsymbol{\epsilon}-\nabla_{\mathbf{h}_t}f_{\boldsymbol{\theta}}(\mathbf{h}_t,t)\right\|_2^2\big].
  \end{equation}
}
    
The parameters $\boldsymbol{\theta}$ can be updated by performing gradient descent steps using $\nabla_{\boldsymbol{\theta}}\|\boldsymbol{\epsilon}-\nabla_{\mathbf{h}_t}f_{\boldsymbol{\theta}}(\mathbf{h}_t,t)\|_{2}^{2}$.
Once the energy-based DM is trained, we fix the network's parameters and use this model to produce channel estimates across different pilot counts, SNR values, and noise distributions.

\CheckRmv{
  \begin{figure}[t]
    \setlength{\abovecaptionskip}{-0.2cm}
    \centering
    \includegraphics[width=3.4in]{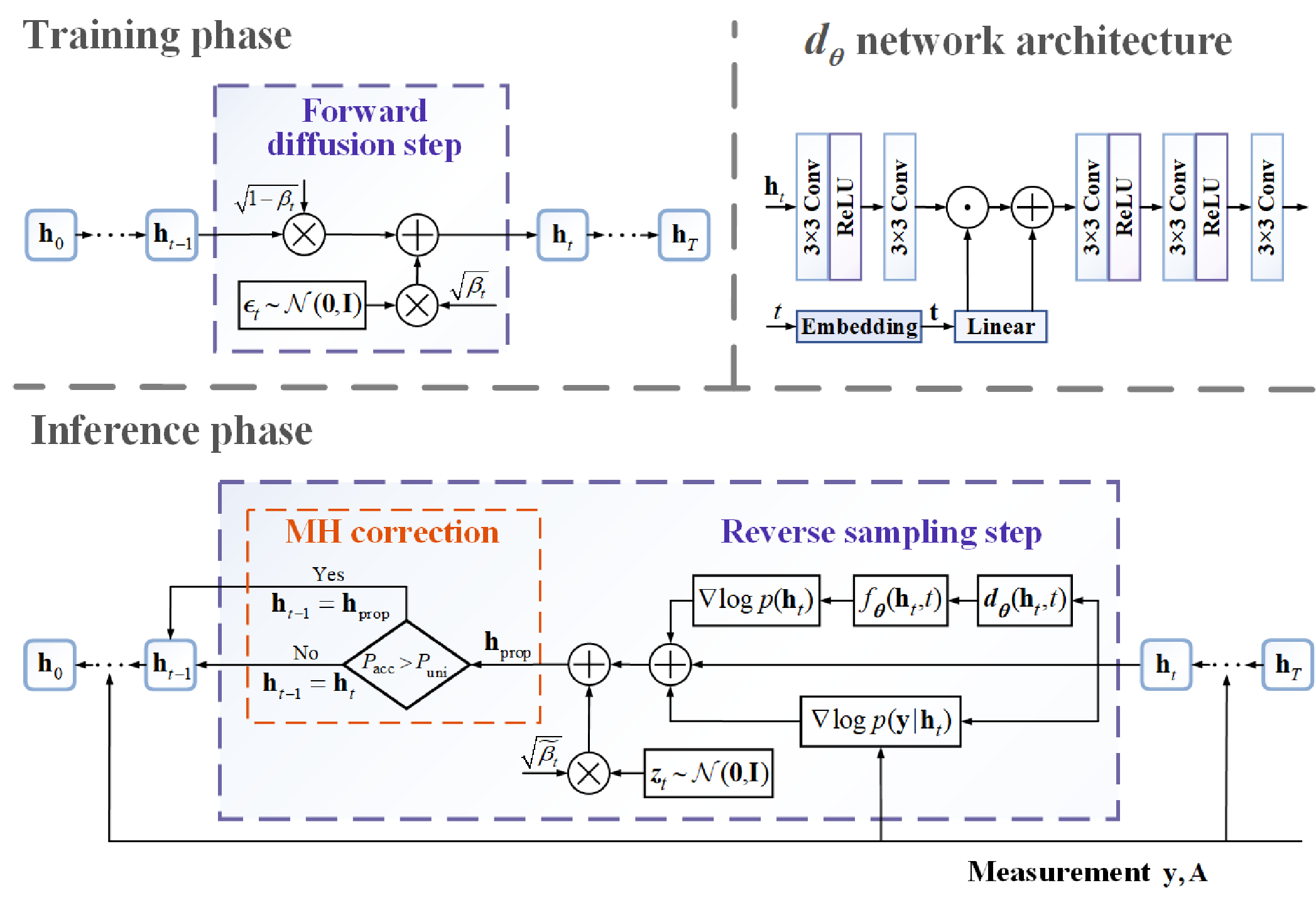}
    \caption{Block diagram of the proposed method.}
    \label{fig:ebm_block}
  \end{figure}
}
\subsection{Inference Process}
The iterative inference process for MIMO channel estimation using the proposed energy-based DM consists of two key steps per iteration:
(1) Constructing a new proposal, and (2) accepting the new proposal subject to the MH acceptance rate.

At each inference time step (from $t=T$ to 1), we first approximate the corresponding prior score by computing the derivative of \eqref{eq:log=prior}
\CheckRmv{
  \begin{equation}
    \nabla_{\mathbf{h}_t}\log p(\mathbf{h}_t)\approx-\frac{1}{\sqrt{1-\bar{\alpha}_t}} \nabla_{\mathbf{h}_t}f_{\boldsymbol{\theta}}(\mathbf{h}_t,t).
    \label{eq:prior}
  \end{equation}
}
Then, with the assumption of an uninformative prior \cite{meng2022diffusion}, the noise-perturbed likelihood can be formulated in the form of a Gaussian distribution given by \cite{zhoudm2025}
\CheckRmv{
  \begin{equation}
    p(\mathbf{y}|\mathbf{h}_t)=\mathcal{N}\left(\mathbf{y};\frac{1}{\sqrt{\bar{\alpha}_t}}\mathbf{Ah}_t,\frac{1-\bar{\alpha}_t}{\bar{\alpha}_t}\mathbf{A}\mathbf{A}^T+\sigma^2\mathbf{I}\right),
    \label{eq:Gaussianlikelihood}
  \end{equation}
}
and thus the gradient of log-likelihood is derived as
\CheckRmv{
\begin{equation}
  \begin{aligned}
     & \nabla_{\mathbf{h}_t}\log p(\mathbf{y}|\mathbf{h}_t) \\
     & =\frac{1}{\sqrt{\bar{\alpha}_t}}\mathbf{A}^T\left(\frac{1-\bar{\alpha}_t}{\bar{\alpha}_t}\mathbf{A}\mathbf{A}^T+\sigma^2\mathbf{I}\right)^{-1}\left(\mathbf{y}-\frac{1}{\sqrt{\bar{\alpha}_t}}\mathbf{A}\mathbf{h}_t\right).
    \label{eq:likelihood}
  \end{aligned}
\end{equation}
}
Furthermore, we employ a gradient scale parameter $s$ in the update rule to balance the contributions of the prior and likelihood \cite{meng2022diffusion}, i.e., the posterior score is formulated as
\CheckRmv{
  \begin{equation}
    \nabla_{\mathbf{h}_t}\log p(\mathbf{h}_t|\mathbf{y})=\nabla_{\mathbf{h}_t}\log p(\mathbf{h}_t)+s\cdot\nabla_{\mathbf{h}_t}\log p(\mathbf{y}|\mathbf{h}_t).
    \label{eq:posteriordecompose}
  \end{equation}
}
Equipped with \eqref{eq:prior}, \eqref{eq:likelihood}, and \eqref{eq:posteriordecompose}, the proposal sample $\mathbf{h}_{\mathrm{prop}}$ at each time step $t$ can be constructed based on \eqref{eq:proposal}.

After obtaining the proposal $\mathbf{h}_{\mathrm{prop}}$, we apply the MH criterion to accept it as $\mathbf{h}_{t-1}$ with the probability given by
\CheckRmv{
  \begin{equation}
    \begin{aligned}
      P_{\mathrm{acc}}&=\min\left(1,\frac{p(\mathbf{h}_{\mathrm{prop}}|\mathbf{y})}{p(\mathbf{h}_t|\mathbf{y})}\cdot\frac{p(\mathbf{h}_t|\mathbf{h}_{\mathrm{prop}})}{p(\mathbf{h}_{\mathrm{prop}}|\mathbf{h}_t)}\right) \\
      &\triangleq\min\left(1,P_{\mathrm{acc}}'\right).
      \label{eq:defi}
    \end{aligned}
  \end{equation}
}
The primary focus here is to derive the expression of $P_{\mathrm{acc}}'$, and we rewrite it in the logarithmic form as follows
\CheckRmv{
  \begin{equation}
    \begin{aligned}
      \log\left(P_{\mathrm{acc}}'\right)  =  & \log p(\mathbf{y}|\mathbf{h}_{\mathrm{prop}})-\log p(\mathbf{y}|\mathbf{h}_t) \\
      & +\log p(\mathbf{h}_{\mathrm{prop}})-\log p(\mathbf{h}_t) \\
      & +\log p(\mathbf{h}_t|\mathbf{h}_{\mathrm{prop}})-\log p(\mathbf{h}_{\mathrm{prop}}|\mathbf{h}_t).
      \label{eq:acc}
    \end{aligned}
  \end{equation}
}
The first line in \eqref{eq:acc} is the difference between the log-likelihoods.
From \eqref{eq:Gaussianlikelihood}, we have
\CheckRmv{
  \begin{equation}
    \begin{aligned}
      &\log p(\mathbf{y}|\mathbf{h}_t)= \\
      &-\frac{1}{2}(\mathbf{y}-\boldsymbol{\mu})^T\left(\frac{1-\bar{\alpha}_t}{\bar{\alpha}_t}\mathbf{A}\mathbf{A}^T+\sigma^2\mathbf{I}\right)^{-1}(\mathbf{y}-\boldsymbol{\mu}) + C',
      \label{eq:acc1}
    \end{aligned}
  \end{equation}
}
where $\boldsymbol{\mu}=\frac{1}{\sqrt{\bar{\alpha}_t}}\mathbf{A}\mathbf{h}_t$ and $C'$ is a constant.
A similar expression for $\log p(\mathbf{y}|\mathbf{h}_{\mathrm{prop}})$ can be derived.
Hence, the constant $C'$ can be cancelled out in the subtraction.
The second line in \eqref{eq:acc} is the difference between the log-priors,
which can be calculated using \eqref{eq:log=prior}.
The third line in \eqref{eq:acc} 
refers to the difference between the logarithm of the reverse and forward transition probabilities. 
According to \eqref{eq:proposal} and \eqref{eq:posteriordecompose}, we can view $p(\mathbf{h}_{\mathrm{prop}}|\mathbf{h}_t)$ as a Gaussian distribution,
\CheckRmv{
  \begin{equation}
      p(\mathbf{h}_{\mathrm{prop}}|\mathbf{h}_t) = \mathcal{N}(\boldsymbol{\mu}_{\boldsymbol{\theta}}(\mathbf{h}_t),\tilde{\beta}_t^2\mathbf{I}), 
      \label{eq:adjustvar}
  \end{equation}
}
where 
\CheckRmv{
  \begin{equation}
    \begin{aligned}
      &\boldsymbol{\mu}_{\boldsymbol{\theta}}(\mathbf{h}_t) =\\
      &\frac{1}{\sqrt{\alpha_t}} \! \left[\mathbf{h}_t \! - \! \frac{1\! - \!\alpha_t}{\sqrt{1\! - \!\bar{\alpha}_t}}\nabla_{\mathbf{h}_t}f_{\boldsymbol{\theta}}(\mathbf{h}_t,t) \!+ \!s(1\! - \!\alpha_t)\nabla_{\mathbf{h}_t}\log p(\mathbf{y}|\mathbf{h}_t)\right].
      \label{eq:acc3}
    \end{aligned}
  \end{equation}
}
The computation for $p(\mathbf{h}_t|\mathbf{h}_{\mathrm{prop}})$ is similar, and the acceptance rate can be derived by combining \eqref{eq:log=prior} and \eqref{eq:defi}-\eqref{eq:acc3}.

The channel estimation scheme using the proposed energy-based DM is outlined in \algref{alg:inference} and illustrated in the lower half of \figref{fig:ebm_block}.

\setlength{\algomargin}{0em} 
\SetAlCapHSkip{0em} 
\CheckRmv{
  \begin{algorithm}[!t]
    \SetAlgoLined
    \SetKwInput{Init}{CU Init.}
    \SetKwInput{Input}{Input}
    \SetKwInput{CU}{CU}
    \SetKwInput{DUone}{Each DU $c\in \mathcal{I}$}
    \SetKwInput{DUtwo}{Each DU $c\in[C]$}
    \SetKwBlock{mcmc}{MCMC sampling stage:}{end}
    \SetKwBlock{nag}{NAG stage:}{end}
    \caption{Channel Estimation via an Energy-Based DM}
    {
    \begingroup

    \Input{$\mathbf{A}$, $\mathbf{y}$, $\sigma^2$, pre-trained denoising network $\boldsymbol{d}_{\boldsymbol{\theta}}$, noise schedule $\{\beta_t\}_{t=1}^T$, gradient scale $s$.}
    {\textbf{Initialize:}  $\mathbf{h}_T \sim \mathcal{N}(\mathbf{0}, \mathbf{I})$ \\ }
    {\textbf{Compute:}   $\alpha_t = 1 - \beta_t$, $\bar{\alpha}_t = \prod_{i=1}^t \alpha_i$, $\tilde{\beta}_{t} =\frac{1-\bar{\alpha}_{t-1}}{1-\bar{\alpha}_{t}}\beta_{t}$. \\}
    \For{$t=T$ \KwTo $1$}
    {
     \setlength\abovedisplayskip{0pt}
     \setlength\belowdisplayskip{0pt}
    {\textbf{Step 1:} Draw a random noise:
    $\mathbf{z}_t\sim\mathcal{N}(\mathbf{0},\mathbf{I})$.\\
    }
    {\textbf{Step 2:} Construct the new proposal:{
        \setlength\abovedisplayskip{0pt}
        \setlength\belowdisplayskip{0pt}\\
        \quad (a) Compute $f_{\boldsymbol{\theta}}(\mathbf{h}_t,t)$ using \eqref{eq:energy}.\\
        \quad (b) Compute $ \boldsymbol{l}_1 = \nabla_{\mathbf{h}_t} \log p(\mathbf{h}_t)$ using \eqref{eq:prior},\\
        \quad \qquad \qquad and $\boldsymbol{l}_2 = \nabla_{\mathbf{h}_t} \log p(\mathbf{y}|\mathbf{h}_t) $ using \eqref{eq:likelihood}.\\
        \quad (c) Compute the posterior score:\\
        \quad \quad  $\nabla_{\mathbf{h}_t}\log p(\mathbf{h}_t|\mathbf{y})=\boldsymbol{l}_1+s\cdot\boldsymbol{l}_2$.\\
        \quad (d) Compute the new proposal using \eqref{eq:proposal}.\\
    }
    }

    {\textbf{Step 3:} Calculate MH-acceptance probability:{
        \setlength\abovedisplayskip{0pt}
        \setlength\belowdisplayskip{0pt}\\
        \quad (a) Compute $\log\left(P_{\mathrm{acc}}\right)$ via \eqref{eq:log=prior}, \eqref{eq:defi}-\eqref{eq:acc3}.\\
        \quad (b) Sample from uniform distribution $P_{\mathrm{uni}}\sim\mathcal{U}(0,1)$.

    }
    }

    {\textbf{Step 4:} MH corrections:{
        \setlength\abovedisplayskip{0pt}
        \setlength\belowdisplayskip{0pt}\\
        \quad \textbf{If} $\log\left(P_{\mathrm{acc}}\right) > \log\left(P_{\mathrm{uni}}\right)$ \textbf{then} $\mathbf{h}_{t-1} = \mathbf{h}_{\mathrm{prop}}$. \\
        \quad \textbf{else} $\mathbf{h}_{t-1} = \mathbf{h}_t$.
    }
    }\\
    }
    {\textbf{Output:} Estimated channel $\hat{\mathbf{h}}=\mathbf{h}_0$. \\}

     \endgroup
    }
  \label{alg:inference}
  \end{algorithm}
    }

  
  
  
    
    
  

\section{Numerical Results}
\label{sec:simulation}
Consider a massive MIMO system with $(N_{\mathrm{t}}, N_{\mathrm{r}})=(64,16)$. 
We utilize the QuaDRiGa toolbox \cite{jaeckelquadriga2020} to generate the training, testing, and validation datasets, containing 10,000, 100, and 100 channel realizations, respectively. 
The proposed energy-based DM is trained for 500 epochs with a batch size of 128. The Adam optimizer is employed with the learning rate fixed as $10^{-4}$. 
The total number of reverse denoising steps $T$ is set as 100, and we adopt the linear noise schedule $\{\beta_t\}_{t=1}^T$ prescribed in \cite{feslnoiseschedule2025}.
We measure estimation accuracy using the normalized mean squared error (NMSE), defined as $\mathrm{NMSE}=\mathbb{E}\{{\|\hat{\mathbf{h}}-\mathbf{h}\|_2^2}/{\|\mathbf{h}\|_2^2}\}$.

We compare the proposed method with the following baselines:
\textbf{Regularized least square (RLS)}: A linear estimator that estimates the channel using the RLS algorithm, given by $\hat{\mathbf{h}}_\text{RLS}=(\mathbf{A}^T\mathbf{A}+\sigma^2\mathbf{I})^{-1}\mathbf{A}^T\mathbf{y}$. 
\textbf{LMMSE}: A linear minimum mean square error estimator using all training samples to compute the global sample covariance matrix 
$\mathbf{C}_h$,
and calculating the estimate as $\hat{\mathbf{h}}_\text{LMMSE}=\mathbf{C}_h\mathbf{A}^T(\mathbf{A}^T\mathbf{C}_h\mathbf{A}+\sigma^2\mathbf{I})^{-1}\mathbf{y}$ \cite{signalprocessing}.
\textbf{Low-complexity DM \cite{fesllowcommplexity2024}}: A low-complexity estimator based on the DM, which requires orthogonal and full pilot observations for initialization.
\textbf{SGM}: Leverage SGMs as the channel prior and perform posterior sampling using annealed Langevin dynamics \cite{arvintescore-based2023}.
\textbf{L-DAMP}: A supervised, data-driven channel estimator that incorporates a denoising CNN with the unfolded approximate message passing (AMP) method. Separate L-DAMP models are trained for each value of pilot density and SNR \cite{heldamp2018}.
\textbf{DM in \cite{zhoudm2025}}: An estimator with the same architecture and hyperparameters as the proposed energy-based DM, but without MH corrections, meaning that all new proposals are accepted.

\CheckRmv{
  \begin{figure}[t]
    \setlength{\abovecaptionskip}{-0.2cm}
    \centering
    \includegraphics[width=3in]{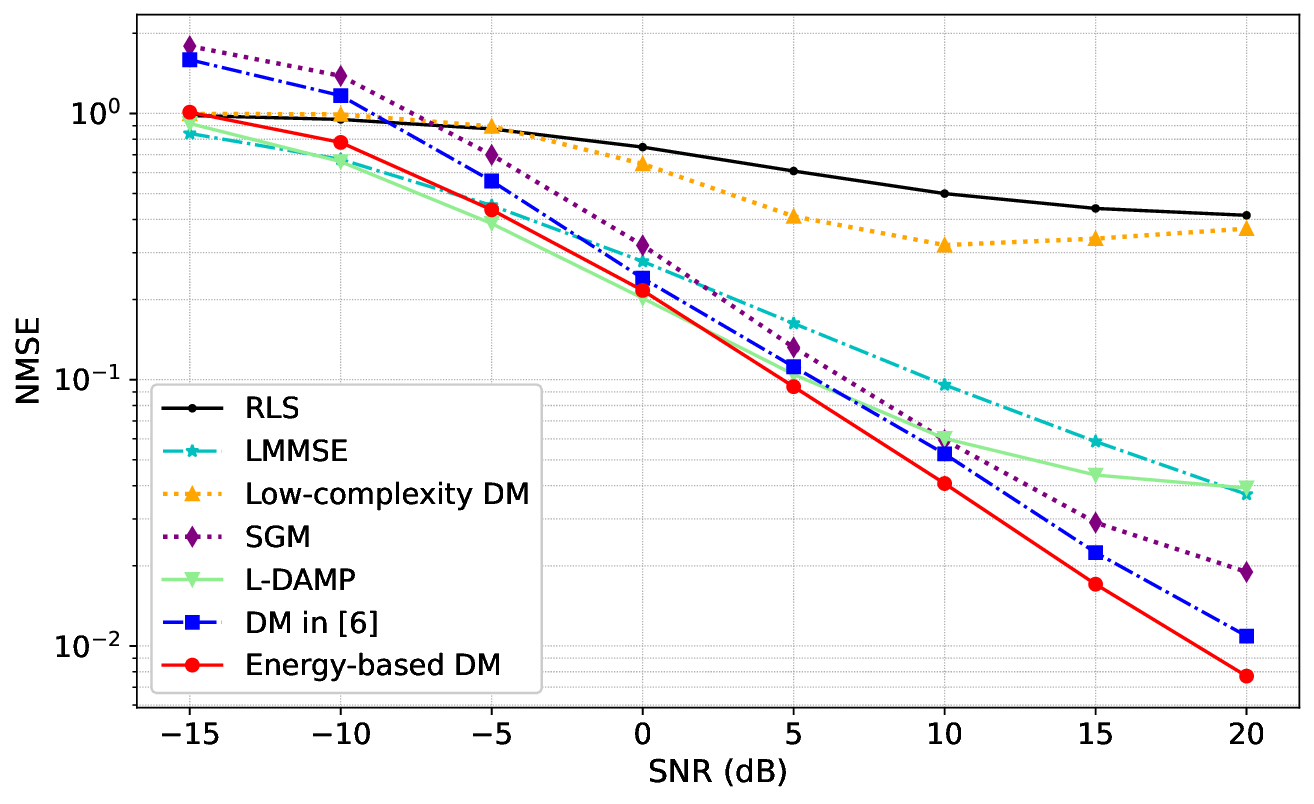}
    \caption{NMSE performance when $\alpha = 0.6$.}
    \label{fig:performance0.6}
  \end{figure}
}

\figref{fig:performance0.6} illustrates the NMSE performance as a function of the SNR when the pilot count is $N_{\text{p}}=38$ $(\alpha=N_{\text{p}}/N_{\text{t}}\approx0.6)$.
From the figure, the proposed energy-based DM method demonstrates exceptional performance across all SNR regions. At low SNRs, it is competitive with the best-performing L-DAMP and LMMSE, while at high SNRs, it outperforms all baseline methods by a noticeable margin. In particular, the proposed method shows clear gains over DM in \cite{zhoudm2025}, verifying the effectiveness of incorporating the MH adjustment through energy-based formulation. 

\CheckRmv{
  \begin{figure}[t]
    \setlength{\abovecaptionskip}{-0.2cm}
    \centering
    \includegraphics[width=3in]{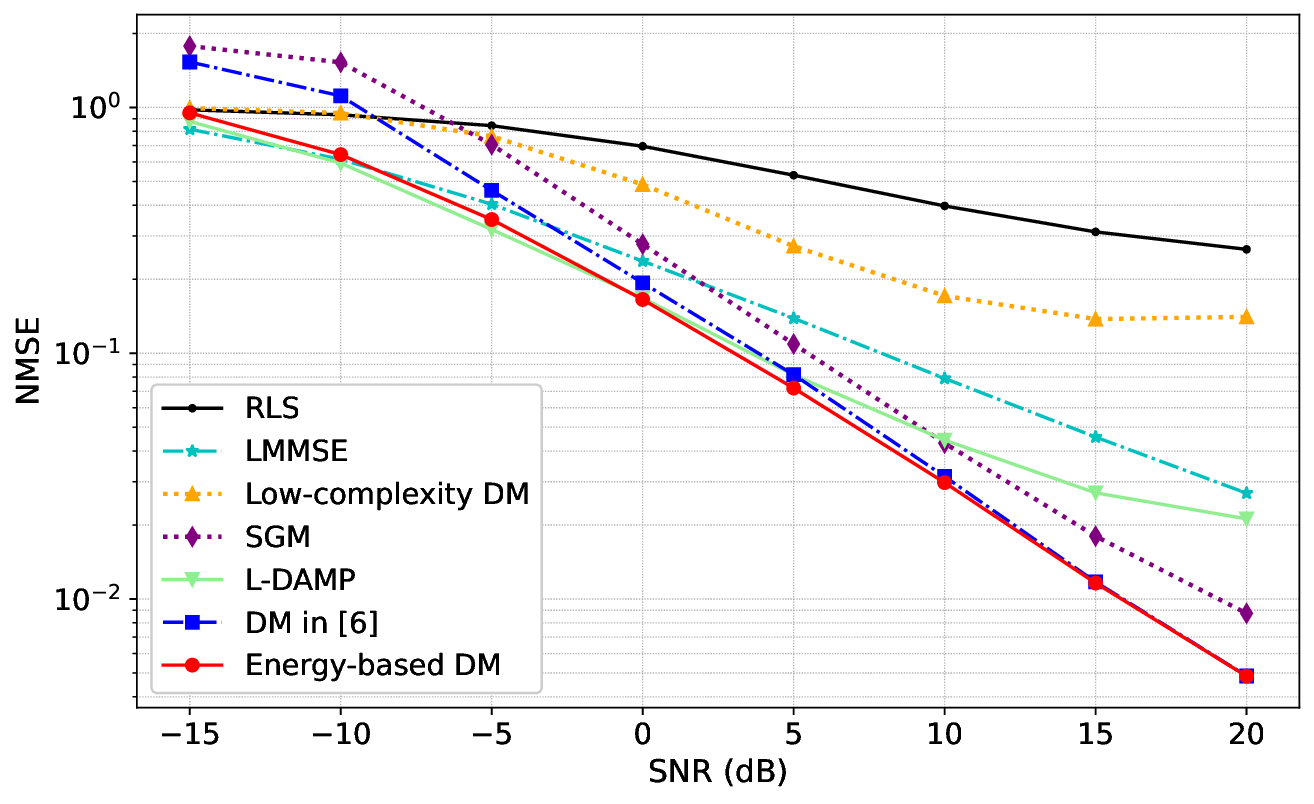}
    \caption{NMSE performance when $\alpha = 0.8$.
    }
    \label{fig:performance0.8}
  \end{figure}
}

\figref{fig:performance0.8} illustrates the NMSE performance when the pilot count is $N_{\text{p}}=51$ $(\alpha=N_{\text{p}}/N_{\text{t}}\approx0.8)$.
Similar observations of the performance comparison can be made as in Fig.~\ref{fig:performance0.6}. 
It is worthy of noting that in this higher pilot density scenario,
the performance gap between the proposed algorithm and DM in \cite{zhoudm2025} narrows as SNR increases, resulting in nearly identical NMSE at high SNRs, although the gains at low SNRs are still substantial.
We can thus conclude that MH corrections provide greater benefits in scenarios with low pilot density and low SNRs, where the channel estimation task becomes particularly challenging. 
\CheckRmv{
  \begin{figure}[t]
    \setlength{\abovecaptionskip}{-0.2cm}
    \centering
    \includegraphics[width=3in]{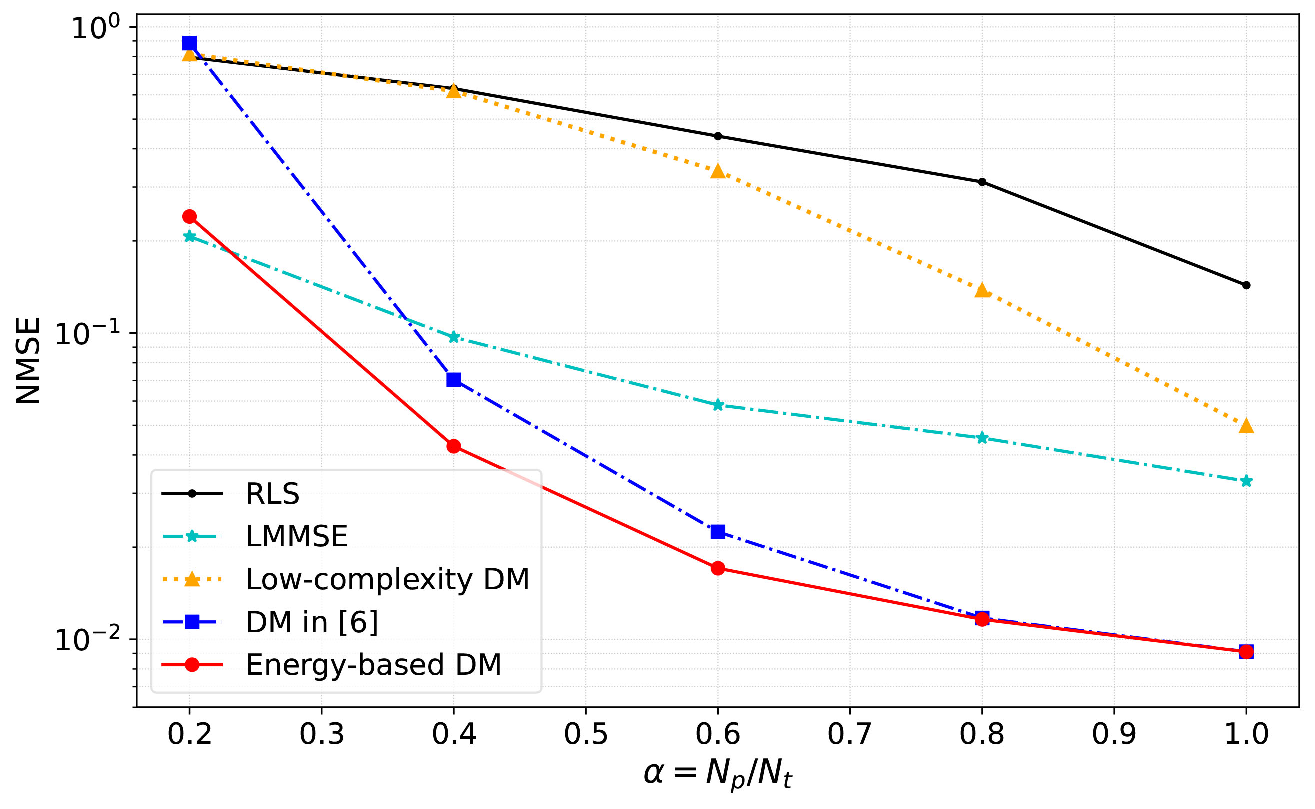}
    \caption{NMSE versus pilot density with SNR = 15 dB.}
    \label{fig:alphavar}
  \end{figure}
}

\figref{fig:alphavar} presents the NMSE performance as a function of the pilot density $\alpha$ with the SNR fixed at 15 dB.
As shown in the figure, the energy-based DM outperforms DM in \cite{zhoudm2025} and other baseline algorithms across the pilot density setups. 
Moreover, compared with DM in \cite{zhoudm2025}, the proposed algorithm achieves significant gains in low pilot density regions. 
These results further underscore the effectiveness of the proposed energy-based DM in improving channel estimation accuracy, particularly in situations with limited pilot overhead.

\CheckRmv{
  \begin{table}[t]
    \centering
    \caption{Complexity And Latency Comparison}
    \label{tab:complexity}
    \renewcommand{\arraystretch}{1.05}
    \begin{tabular}{|c|c|c|}
    \hline
    \textbf{Algorithm} & \textbf{FLOPs ($\times 10^8 $)} & \textbf{Runtime (ms)}\\
    \hline
    RLS & 1.6565 &0.5573 \\
    \hline
    LMMSE &9.2561 &282.8270  \\
    \hline
    Low-complexity DM & 13.8167 &0.4294  \\
    \hline
    SGM &56987.7965 &1016.4458  \\
    \hline
    L-DAMP & 13.8320 & 2.4494 \\
    \hline
    DM in \cite{zhoudm2025} & 224.2998 & 8.0646 \\
    \hline
    Energy-based DM & 224.3023 & 8.1439\\
    \hline
    \end{tabular}
    \label{tab:complexities}
  \end{table}
}

\tabref{tab:complexities} presents the floating point operation counts (FLOPs) and runtime of different estimators under a single channel realization.
These metrics are evaluated on a machine with an NVIDIA RTX 4090 GPU and an Intel Xeon Platinum 8352V CPU.
The proposed algorithm demonstrates significantly lower computational costs than SGM in terms of both runtime and FLOPs. The runtime is also maintained at manageable levels compared with all baselines. Additionally, the complexity of DM in \cite{zhoudm2025} and energy-based DM is nearly identical, while incorporating MH corrections substantially improves the estimation accuracy.
This demonstrates that the proposed energy-based DM achieves a favorable trade-off between computational efficiency and performance.

\section{Conclusion}

We have proposed a novel massive MIMO channel estimation approach, leveraging the energy-parameterized DM with a lightweight network architecture.
By introducing an energy function, we explicitly estimate the unnormalized log-prior and incorporate MH corrections during the inference phase. 
Simulation results demonstrate that the proposed algorithm exhibits significant superiority in terms of NMSE over state-of-the-art estimators.
Moreover, the energy-based DM shows remarkable advantages in scenarios with limited pilot overhead, improving estimation accuracy while maintaining moderate computational complexity.

\vspace{-0.1cm}




\ifCLASSOPTIONcaptionsoff
  \newpage
\fi



\end{document}